# The Influence of Filaments in the Private Flux Region on Divertor Particle and Power Deposition

J. R. Harrison[1], G. M. Fishpool[1], A. J. Thornton[1], N. R. Walkden[2] and the MAST team[1]

[1] CCFE, Culham Science Centre, Abingdon, Oxon, OX14 3DB, UK

[2] York Plasma Institute, Department of Physics, University of York, Heslington, York, YO10 5DD, UK

**Abstract**

The transport of particles via intermittent filamentary structures in the private flux region of plasmas in the MAST tokamak has been investigated using a fast framing camera recording visible light emission from the volume of the lower divertor, as well as Langmuir probes and IR thermography monitoring particle and power fluxes to plasma-facing surfaces in the divertor. The visible camera data suggests that, in the divertor volume, fluctuations in light emission above the X-point are strongest in the scrape-off layer (SOL). Conversely, in the region below the X-point, it is found that these fluctuations are strongest in the private flux region (PFR) of the inner divertor leg. Detailed analysis of the appearance of these filaments in the camera data suggests that they are approximately circular, around 1-2cm in diameter. The most probable toroidal mode number is between 2 and 3. These filaments eject plasma deeper into the private flux region, sometimes by the production of secondary filaments, moving at a speed of 0.5-1.0km/s. Probe measurements at the inner divertor target suggest that the fluctuations in the particle flux to the inner target are strongest in the private flux region, and that the amplitude and distribution of these fluctuations are insensitive to the electron density of the core plasma, auxiliary heating and whether the plasma is single-null or double-null. It is found that the e-folding width of the time-average particle flux in the PFR decreases with increasing plasma current, but the fluctuations are unchanged. At the outer divertor target, the fluctuations in particle and power fluxes are strongest in the SOL.

# 1. Introduction

The fluxes of power and particles to divertor plasma-facing components (PFCs) is a concern in large-scale magnetic confinement fusion devices such as ITER [1, 2] and DEMO [3], as these fluxes can result in damage of the surfaces of PFCs and the sputtering of impurities that can adversely affect the performance of the core plasma. These loads can be reduced by designing divertors with high poloidal flux expansion [4] and surfaces oriented to maximise the wetted area, to ensure that recycled neutrals re-ionize preferentially near the separatrix where these loads are greatest [5] and impurity seeding [6, 7]. In DEMO, reducing these loads to levels tolerable with existing technology and materials is highly challenging [3]. This has motivated the development of alternative divertor magnetic configurations such as the snowflake [8] and Super-X [9] that aim to increase the wetted area and increase the potential for power and particle exhaust through radiation. It has also been speculated that these configurations could result in increased cross-field transport [9, 10], thereby further increasing the wetted area. At present, the transport of particles and power against field lines cannot be accurately modelled from first principles, and is the subject of significant experimental, theory and modelling efforts [11, 12, 13, 14, 15, 16, 17]. In particular, experimental and theoretical observations of filamentary transport in the vicinity of the X-point are presented in [18, 19, 20, 21, 22, 23].

The divertor of the MAST tokamak is well-suited to studying the processes that govern the cross-field transport of particles, with a geometrically open design [24], allowing significant access for diagnostics imaging light emission from the divertor volume. Furthermore, the divertor target plates are instrumented with Langmuir probes, which can measure the particle flux to the inner and outer targets with a typical spatial resolution of 5-10 mm. This work builds on the recent observation of intermittent blobs of plasma, extended along magnetic field lines, referred to as filaments, in the private flux region of MAST [25], by presenting detailed analysis of the light emitted by these filaments to deduce their toroidal mode number, size and radial propagation velocity. These filaments appear brightest in the private flux region of the inner leg, and their influence on the distribution of particle fluxes to the inner target plate is inferred by ascertaining the sensitivity of the particle flux distribution, and the fluctuations in these fluxes due to filaments, to conditions in the core plasma.

In this paper, details of the diagnostics used in this study are given in section 2. The spatial distribution of fluctuations in light emission in the divertor and the relative contribution of these fluctuations to the time-average light emitted in the inner divertor leg is discussed in section 3. Measurements of visible light emission from intermittent "filaments" of plasma in the volume of the private flux region of the inner divertor leg are presented in section 4, including estimates of the toroidal mode number, size, radial propagation velocity. The influence of these filaments on the distribution of the particle flux to the inner strike point is discussed in section 5. These observations are compared with measurements of particle and power fluxes to the outer strike point in section 6. Finally, a discussion of how these observations contribute to our understanding of cross-field transport in the private flux region is given in section 7 followed by a summary in section 8.

## 2. Experiment set-up

A poloidal cross-section through the MAST vessel, together with a typical plasma shape and diagnostic layout is shown in Figure 1. The open vessel and divertor geometry allows for excellent viewing access for imaging diagnostics viewing the main chamber or divertor regions in MAST. Most discharges in this study are in a lower single-null configuration, as it is possible to view the plasma from several cm above the X-point to the inner and outer strike points in a single frame, with minimal vignetting from poloidal field coils or supporting infrastructure. More details of the camera field of view are given in [25]. The principal diagnostic used in this study is a Photron SA1.1 fast framing camera, with a CMOS detector, operated by reading out a 160x192 region within the full 1024x1024 sensor, to allow the whole lower divertor volume to be imaged at a frame rate of 120kHz. The camera was unfiltered to maximise the light throughput of the imaging optics, which was made up of three f/2 lenses to provide a $28^0$ horizontal and $33^0$ vertical field of view and to act as relay lenses, to increase the distance between the camera and MAST, thereby reducing electromagnetic interference experienced by the camera.

Arrays of flush-mounted Langmuir probes were used to monitor the ion current to the divertor target plates [26], which have 5mm and 10mm spacing at the inner and outer divertor targets respectively. The bias voltage applied to the probes was swept, and groups of 16 probes were multipliexed to a single power supply, as this is the default operating mode. The ion current to the target was extracted from the swept profiles by isolating the ion saturation region of the I-V characteristic. The current induced by capacitance effects in the signal cables was corrected for by applying the sweeping waveform to the probes when no plasma is present and measuring the current drawn, which is a product of the cable capacitance and the known rate of change of the bias voltage. Approximately 22% of the probe bias waveform, of 65µs duration, was used in the ion current measurements. The probe bias voltage and current drawn were sampled at 1MHz. Radial profiles of the ion current were acquired every 1.04ms.

The temperature of the divertor plasma-facing surfaces was measured using IR cameras [27, 28]. This information was used to estimate the heat flux to the surface of the lower outer divertor target at 2mm spatial resolution and 0.8ms time resolution.

The results presented are from a series of well-diagnosed MAST plasmas where scans in core plasma parameters were carried out, to assess the sensitivity of properties of filaments in the PFR (size, velocity) and the time-averaged particle and power fluxes to the divertor target plates. Scans of core electron density, plasma current, auxiliary heating power and magnetic topology (single and double null) were carried out, over the ranges listed in Table 1.

**3. Fluctuations in Light Emission in the Divertor**

On the scale of a single pixel, if a plasma is free of transients, the time-varying component of the signal on timescales shorter than several kHz is made up of contributions from filaments passing through the pixel sightline, and from electrical noise. Assuming the contribution from electrical noise is negligible, the standard deviation of the signal from each pixel can be used as a measure of the amplitude of the time-varying component of the camera signal. This assumption is justified as the peak signal:noise ratio of the camera data is approximately 1000:1. The standard deviation of camera data taken during 8ms of a H-mode period between ELMs is shown in Figure 2, where fluctuations are strongest in the SOL above the X-point, the PFR of the inner leg and at the divertor strike points. The shape of the fluctuating region in the inner leg PFR conforms well with equilibrium flux surfaces, extending to $\psi_n = 0.98$.

The variation in the time-averaged camera signal along the equilibrium separatrix along the inner divertor leg is shown in Figure 3. This time-average signal is compared with the fluctuation amplitude, calculated as the integral of the camera signal on a given pixel with respect to time divided by the time period of the integration. It can be seen that the fluctuation amplitude peaks near the inner strike point, at approximately 10% of the mean value, and decreases away from the strike point. This decrease in camera signal is consistent with a commensurate decrease in the neutral density, estimated using interpretive OSM-EIRENE simulations [30].

## 4. Properties of Filaments in the Private Flux Region

Images of the filaments in the private flux region were analysed to estimate their basic properties: toroidal mode number, major and minor diameter, and radial propagation velocity. This analysis is restricted to a region in the image in the private flux region of the inner divertor leg, where they are brightest. Although a correlation analysis has shown that these filaments extend along field lines to the outer divertor leg, their brightness is significantly reduced, and there is an additional significant contribution to the camera signal in this region from the separatrix and scrape-off layer of the outer divertor leg.

Unless otherwise stated, the analysis presented in this section used images where a background subtraction technique has been applied, that calculates the minimum signal for a given pixel over ±20 frames (spanning 328μs in total), which is subtracted from the frame of interest [31]. The camera viewing geometry (location, orientation) and imaging properties (effective focal length, distortion) were deduced by fitting the locations of unique structures in the raw images with their known 3D locations. This information was input into a ray-tracing code to deduce the 3D tangency position of all pixels in the images. The positions of equilibrium flux surfaces at the tangency positions were estimated by interpolating the output from the EFIT equilibrium reconstruction code [32].

**4.1 Toroidal Mode Number**

The toroidal mode number of filaments in the private flux region was deduced by counting the number of peaks in the image brightness in a section of an equilibrium flux contour in the private flux region ($\psi_n$ ~0.992). Unlike the main chamber SOL, the high toroidal and low poloidal magnetic field components in the PFR result in field lines crossing the camera tangency plane two or three times, which can result in spuriously high toroidal mode numbers if this is not taken into account. The region of interest used in this analysis spans one toroidal rotation of a field line starting on the contour near the X-point, as shown in Figure 4.

The results of this analysis carried out in the L-mode and H-mode periods of the same discharge (shot number 29564) is shown in Figure 5. It is clear that the filaments are not well described by a single toroidal mode number, and in general span a range from 1 to 5. Similar observations have been made of filaments in the main chamber SOL [11], although the quoted mode numbers are higher, due to the relatively higher poloidal field in that region. It is also clear that, in this case, the most probable toroidal mode number is higher in H-mode compared with L-mode.

**4.2 Filament Size**

The filaments appear in the camera images to be approximately elliptical, therefore in this section measurements of the size of the filaments are expressed in terms of major and minor widths. A sub-set of the image in the private flux region was used in the analysis, bounded by $\psi_n = 0.95$ and $\psi_n = 1.0$ flux contours between the X-point and the inner strike point. A blob detection algorithm was applied to the pixels within this region of interest, to estimate the centre of each filament and the direction of the major and minor axes. An example region of interest and output from the detection algorithm is shown in Figure 6. The major and minor axes output from the blob detection algorithm were used to take profiles of the image brightness along these lines. The minor width of the filaments was calculated by calculating the full-width at half-maximum (FWHM) of the image profile using interpolation and by fitting a Gaussian curve to the image data. Both techniques were found to give comparable results within ~6mm, the distance between adjacent pixels at the tangency plane. The results from the Gaussian fitting procedure are presented here, as it was found that this technique is generally more robust to noise and occasional variations in signal thought to be due to neutrons generated by neutral beam injection. The results of this analysis are shown in Figure 7, where the analysis was carried out in the inter-ELM H-mode period of shot 29264 ($I_p$ = 620kA, $n_{e,core}$ = $4\times10^{19}$m$^{-3}$ and $P_{NBI}$ = 1.2MW). No significant difference between L-mode and H-mode filaments were found in terms of their minor width.

A different analysis technique was required to estimate the major width of the filaments. As indicated in Figure 6, the signal variation along the major axis is too broad to be accurately described by a Gaussian function. Furthermore, the direction of the major axis is well aligned with the magnetic field, so the measurement is significantly influenced by line integration effects. The effect of line integration on the apparent size of object whose light emissivity varied as a function of radius was investigated using a simple model, where the light emissivity was modelled as a Gaussian function as a function of radius and did not vary with toroidal position (see Figure 8). The light emissivity was line integrated using a procedure analogous to the Abel transform [33] to estimate the line integrated signal from a given emissivity function. It was found that the FWHM of the simulated line integrated data is approximately twice that of the emissivity function, and is robust over a broad range of emissivity function widths, spanning the range of detected filament sizes. Interpolation was used to estimate the major width in a procedure analogous to that used to estimate the minor width. To account for line integration effects the measurements were scaled by a factor of 2.05 calculated using simple forward modelling. The results of this analysis are shown in Figure 9. The most probable minor and major widths are comparable, suggesting that the filaments are approximately circular with an average width of ~1-2cm. This filament size is much larger than the ion Larmor radius, which is approximately 1mm, and the 6mm spacing between pixels at the tangency position. The broader distribution in major width of the filaments could be due to magnetic shear as the filaments move toward the inner divertor target, which can distort notionally circular filaments into ellipses [34].

### 4.3 Radial Propagation

As noted in [25], the filaments appear to move along the inner leg toward the inner strike point, either poloidally or toroidally, at approximately 1-2km/s or 10-20krad/s respectively. It is also observed that, as filaments execute this motion, they eject plasma away from the separatrix. The motion of one such "secondary" filament is shown in Figure 10, and does not appear to be well aligned to either the curvature vector or $-\nabla$ . In general, it is observed that the motion of these filaments is predominantly in the radial direction deeper into the PFR, with a velocity ranging from 0.5-1.5km/s. The process of filaments ejecting plasma radially in the private flux region is a frequent occurrence in the visible camera data.

### 5. Fluctuations at the Inner Strike Point

The primary diagnostics for monitoring plasma interaction with the inner target plates in this study are the visible imaging camera, and arrays of flush-mounted Langmuir probes embedded in the divertor tiles. The camera and probes provide profiles of light emission from, and the flux of ions to the inner target plate. The fluctuating light and ion current signals are decomposed into a mean average and standard deviation to assess the influence of fluctuations on the average profiles. A typical profile of the ion flux to the inner target is shown in Figure 11. It is commonly observed in these profiles at the inner strike point that the peak of the standard deviation is in the private flux region and offset from the peak in the mean value.

A comparison of the normalised profiles measured using the camera and Langmuir probes is shown in Figure 12. In the camera data, the contribution to the measured signal from the outer divertor leg was removed by subtracting the baseline from the profile in the private flux region. This simple background subtraction technique is adequate in the private flux region as a small region of the outer leg, approximately 5cm in poloidal distance along the leg near the x-point, contributes to the signal in this region, which is short compared with the gradient scale lengths of $D_\alpha$ emission. The profiles are in good agreement in terms of the profile shape in the private flux region, the focus of this study. The level of agreement in the SOL is reduced by an additional contribution to the measured camera signal from the SOL of the inner divertor leg.

The influence of filaments in the private flux region on the width of the average profiles was investigated by measuring profiles in separate scans of core plasma density, auxiliary heating power, magnetic topology and plasma current. With the exception of the core density ramp, which was conducted in a single shot, each scan was conducted in repeated shots such that all other plasma properties were as close to other plasmas in the scan as possible. This was achieved within an accuracy of ~10%. In all scans the inner divertor leg remained attached throughout. The range over which the core plasma parameters were scanned is shown in Table 1. It was found that profiles of both the average and standard deviation of the light emission from the private flux region and the ion flux were unaffected by scans in core density, auxiliary heating and magnetic topology within the experimental uncertainties in the measurements. However, it was found that the decay length of the time-averaged profiles decreased with increasing plasma current, where $\lambda_{PFR}(\Gamma_n) = 0.015$ in the $I_p = 400$kA shot and $\lambda_{PFR}(\Gamma_n) = 0.010$ in the $I_p = 650$kA shot, as shown in Figure 13. The reduction of $\lambda_{PFR}$ with increasing $I_p$ is analogous to the commonly observed behaviour in profiles in the SOL of the outer divertor leg on several devices [35, 36, 28, 37]. Despite the reduction of $\lambda_{PFR}(\Gamma_n)$ of the average profiles, the fluctuating components of the profiles are largely unaffected.

## 6. Fluctuations at the Outer Strike Point

Radial profiles of the ion current and heat flux to the outer divertor were measured using Langmuir probes and an IR camera respectively. Typical profiles from the Langmuir probes are shown in Figure 14, where in both L-mode and H-mode, the strongest fluctuations are measured near the separatrix and in the SOL, with a clear transition between the "near" and "far" SOL observed in the amplitude of the fluctuations. These observations at the outer strike point, where the fluctuations are strongest in the SOL, are clearly very different to those made at the inner strike point, where they are strongest in the PFR. This is in agreement the camera data that showed that the filaments in the PFR are strongest near the inner divertor leg. Furthermore, the profiles of the divertor heat flux deduced by IR camera measurements indicate that the fluctuations are strongest near the separatrix an in the SOL.

One possible explanation for this imbalance is the role of magnetic shear on the filaments. If the filaments are generated in the PFR of the inner leg approximately circular in cross-section, field line tracing calculations [34] shown in Figure 15 suggest that they would be strongly sheared by the X-point as the filament extends along field lines toward the outer divertor target.

## 7. Discussion

The experimental observation of filaments in the private flux region is promising, both as an additional cross-field transport mechanism (in addition to steady-state diffusion and cross-field drifts) in the divertor to broaden the region over which particles and power are deposited to plasma-facing surfaces, and to test models of filament generation and transport that are normally applied to the region above the X-point. The analysis presented in [25] suggests a lack of correlation between filaments in the PFR and their counterparts in the SOL, perhaps indicating the filaments in the PFR are generated by local instabilities.

These filaments eject plasma deeper into the PFR, manifesting as fluctuations in signals measured at the divertor target plates. The amplitude of these fluctuations at the inner divertor, where they are strongest, peak at 10-15% of the mean value and appear to be independent on the e-folding width of the PFR measured at the inner target. These observations suggest that these filaments, on the length and timescales resolvable by the fast cameras and Langmuir probes, could be an important mechanism for particle transport in the PFR. The reduction of the e-folding width of the particle flux profile in the private flux region with increasing plasma current does not result in a significant change in the amplitude and distribution of fluctuations at the inner strike point. This suggests that the filaments, which manifest as fluctuations in measurements at the divertor target, do not strongly influence the width of the time-average profiles. This could indicate that there are stronger cross-field transport mechanisms at smaller length scales than the diagnostics used in this study are capable of resolving, or that other mechanisms such as magnetic drifts play a strong role, as suggested in [38]. The existence of these filaments is also a consideration for accurately modelling the plasma properties in the private flux region, which can influence the attenuation of neutrals by the plasma in the divertor region, introduced either through recycling and/or fuelling, and the plasma interaction with baffling structures in the divertor.

A significant open question is the role, if any, of these filaments in divertor detachment. It is possible that these filaments play a role in the "fluctuating detached state" observed at AUG [39], as a power spectrum of camera data from a pixel in the private flux region, shown in Figure 16, suggests that the frequencies of these phenomena are comparable, around 5-10kHz. Any connection between these observations could result in a greater understanding of cross-field transport in detached plasmas and the interpretation of experimental measurements, which often average over fluctuations occurring on short timescales.

The similarities and differences between the properties and motion of the filaments in the PFR and those in the SOL could be an important benchmark for models of filament formation and transport. One striking feature of the motion of filaments in the PFR is that their radial motion is mostly associated with the ejection of secondary structures deeper into the PFR, as opposed to the situation in the SOL, where it is generally observed that all L-mode and inter-ELM filaments propagate radially. As the poloidal extent of the private flux region is often much smaller than the scrape-off layer, the camera data from MAST captures light emission from the whole private flux region, and is thus able to fully capture the motion of the filaments throughout their lifecycle, which is not often possible for filaments in the SOL, and could be a strong constraint for models describing edge turbulence.

## 8. Summary

Data from a fast visible imaging camera, Langmuir probes and IR thermography have been used to characterise filaments in the private flux region of MAST and to estimate their effect on time-averaged profiles at the divertor targets. It has been found that the filaments are most apparent in the private flux region of the inner leg, have a toroidal mode number ranging from 1-4, are approximately circular, with a width of approximately 1-2cm (~10-20 ion Larmor radii) and propagate radially at approximately 0.5-1.5km/s. These filaments manifest as fluctuations at the divertor targets, are strongest at the inner target, at around 10-15% of the mean value, and are weaker at the outer target, where the fluctuations are strongest in the scrape-off layer.

**Acknowledgements**

This work has been carried out within the framework of the EUROfusion Consortium and has received funding from the Euratom research and training programme 2014-2018 under grant agreement No 633053. The views and opinions expressed herein do not necessarily reflect those of the European Commission. To obtain further information on the data and models underlying this paper please contact PublicationsManager@ccfe.ac.uk. The views and opinions expressed herein do not necessarily reflect those of the European Commission. We acknowledge the use made of the Photron SA1.1 which was borrowed from the EPSRC (Engineering and Physical Sciences Research Council) Engineering Instrument Pool.

# References

[1] A. Loarte, B. Lipschultz, A.S. Kukushkin, G.F. Matthews, P.C. Stangeby, N. Asakura, G.F. Counsell, G. Federici, A. Kallenbach, K. Krieger, et al., Nucl. Fusion **47**, S203-S263 (2007)

[2] M. Shimada, D.J. Campbell, V. Mukhovatov, M. Fujiwara, N. Kirneva, K. Lackner, M. Nagami, V.D. Pustovitov, N. Uckan, J. Wesley, et al., Nucl. Fusion **47**, S1-S17 (2007)

[3] R. P. Wenninger, M. Bernert, T. Eich, E. Fable, G. Federici, A. Kallenbach, A. Loarte, C. Lowry, D. McDonald, R. Neu, T. Pütterich, P. Schneider, B. Sieglin, G. Strohmayer, F. Reimold, M. Wischmeier, Nucl. Fusion **54**, 114003 (2014)

[4] M. Kotschenreuther, P. Valanju, J. Wiley, T. Rognlein, S. Mahajan, M. Pekker, in Proceedings of 2004 IAEA Fusion Energy Conference, Vilamoura, Portugal, 1–6 Nov 2004, Paper IC/P6-43 , available at http://www-naweb.iaea.org/napc/physics/fec/fec2004/papers/ic_p6-43.pdf.

[5] A. Loarte, Plasma Phys. Control. Fusion **43**, R183-R224 (2001)

[6] A. Kallenbach, M. Bernert, R. Dux, L. Casali, T. Eich, L. Giannone, A. Herrmann, R. McDermott, A. Mlynek, H. W. Müller, et al., Plasma Phys. Control. Fusion **55**, 124041 (2013)

[7] M. E. Fenstermacher, J. Boedo, R. C. Isler, A. W. Leonard, G. D. Porter, D. G. Whyte, R. D. Wood, S. L. Allen, N. H. Brooks, R. Colchin, et al., Plasma Phys. Control. Fusion **41**, A345-A355 (1999)

[8] D. D. Ryutov, Phys. Plasmas **14**, 064502 (2007)

[9] P. M. Valanju, M. Kotschenreuther, S. M. Mahajan, Fusion Engineering and Design **85**, 46-52 (2010)

[10] D. D. Ryutov, R. H. Cohen. T. D. Rognlien, M. V. Umansky, Contrib. Plasma Physics **52** 539–43 (2012)

[11] D. A. D'Ippolito, J. R. Myra, S. J. Zweben, Phys. Plasmas **18**, 060501 (2011)

[12] J. A. Boedo, J. Nucl. Mater. **390-391**, 29-37 (2009)

[13] O. E. Garcia, S. M. Fritzner, R. Kube, I. Cziegler, B. LaBombard, J. L. Terry, Phys. Plasmas **20**, 055901 (2013)

[14] M. Ko an, H.W. Müller, B. Nold, T. Lunt, J. Adámek, S.Y. Allan, M. BernerT, G.D. Conway, P. de Marné, T. Eich, et al., Nucl. Fusion **53**, 073047 (2013)

[15] V. Naulin, J. Nucl. Mater. **363-365**, 24-31 (2007)

[16] B. D. Dudson, N. Ben Ayed, A. Kirk, H. R. Wilson, G. Counsell, X. Xu, M. Umansky, P. B. Snyder, B. Lloyd, Plasma Phys. Control. Fusion **50**, 124012 (2008)

[17] N. Ben Ayed, A. Kirk, B. Dudson, S. Tallents, R. G. L. Vann, H. R. Wilson, Plasma Phys. Control. Fusion **51**, 035016 (2009)

[18] J. L. Terry, S. J. Zweben, M. V. Umansky, I. Cziegler, O. Grulke, B. LaBombard, D.P. Stotler., J. Nucl. Mater. **390-391**, 339-342 (2009)

[19] M. V. Umansky, T. D. Rognlien, X. Q. Xu, J. Nucl. Mater. **337-339**, 266-270 (2005)

[20] R. H. Cohen, D. D. Ryutov, Contrib. Plasma Phys. **46**, 678-684 (2006)

[21] D. D. Ryutov, R. H. Cohen, Contrib. Plasma Phys. **48**, 48-57 (2008)

[22] F. Militello, W. Fundamenski, V. Naulin, A. H. Neilsen, J. Nucl. Mater. **438**, S530-S535 (2013)

[23] F. D. Halpern, P. Ricci, B. Labit, I. Furno, S. Jolliet, J. Loizu, A. Mosetto, G. Arnoux, J. P. Gunn, J. Horacek, Nucl. Fusion **53**, 122001 (2013)

[24] A.C. Darke, R.J. Hayward, G.F. Counsell, K. Hawkins, Fusion Engineering and Design **75-79**, 285-289 (2005)

[25] J. R. Harrison, G. M. Fishpool, B. D. Dudson, Submitted to J. Nucl. Mater.

[26] J. W. Ahn, PhD Thesis, Imperial College

[27] G. De Temmerman, E. Delchambre, J. Dowling, A. Kirk, S. Lisgo, P. Tamain, Plasma Phys. Control. Fusion **52**, 095005 (2010)

[28] A. J. Thornton, A. Kirk, Plasma Phys. Control. Fusion **56**, 055008 (2014)

[29] M. Greenwald, J.L. Terry, S.M. Wolfe, S. Ejima, M.G. Bell, S.M. Kaye and G.H. Neilson, Nucl. Fusion **28** 2199–207 (1988)

[30] S. Lisgo, P. Börner, C. Boswell, D. Elder, B. LaBombard, B. Lipschultz, C. S. Pitcher, D. Reiter, P. C. Stangeby, J. L. Terry, S. Wiesen, J. Nucl. Mater. **337-339**, 139-145 (2005)

[31] B. D. Dudson, PhD Thesis

[32] L.L. Lao, H. St. John, R.D. Stambaugh, A.G. Kellman and W. Pfeiffer, Nucl. Fusion **25** 1611-1622 (1985)

[33] I. H. Hutchinson, "Principles of Plasma Diagnostics" second edition, Cambridge University Press (2002)

[34] D. Farina, R. Pozzoli, D. D. Ryutov, Nucl. Fusion **33**, 1315-1317 (1993)

[35] T. Eich, B. Sieglin, A. Scarabosio, A. Herrmann, A. Kallenbach, G. F. Matthews, S. Jachmich, S. Brezinsek, M. Rack, R. J. Goldston, J. Nucl. Mater. **438**, S72-S77 (2013)

[36] C. J. Lasnier, M. A. Makowski, J. A. Boedo, S. L. Allen, N. H. Brooks, D. N. Hill, A. W. Leonard, J. G. Watkins, W. P. West, J. Nucl. Mater. **415**, S353-S356 (2011)

[37] J. R. Harrison, G. M. Fishpool, A. Kirk, J. Nucl. Mater. **438**, S375-S378 (2013)

[38] R. J. Goldston, Nucl. Fusion **52**, 013009 (2012)

[39] S. Potzel, M. Wischmeier, M. Bernert, R. Dux, H. W. Müller, A. Scarabosio, Nucl. Fusion **54**, 013001 (2014)

**Figure Captions**

Figure 1: Poloidal cross-section of the MAST vessel with the last closed flux surface of a typical lower single null configuration shown in red. The field of view of a high-speed camera recording visible light emission from the volume of the lower divertor is shown in blue. The particle and power fluxes to the divertors were monitored using Langmuir probes indicated in green and IR cameras, not shown.

Figure 2: Standard deviation of raw camera data taken during 8ms of a H-mode period between ELMs. The contour of $\psi_n=0.98$ at the tangency plane is shown in yellow and the poloidal field coils and divertor PFCs are overlaid in gray.

Figure 3: Variation in the time-averaged camera signal along the inner divertor leg at the separatrix (black, left axis) compared with an equivalent fluctuation amplitude relative to the time-averaged signal (blue, right axis) where the fluctuations are strongest, at $\psi_n = 0.992$. The decrease in the time-averaged and fluctuating signal away from the strike point is consistent with a decrease in the neutral density, estimated by interpretive OSM-EIRENE simulations (red dashed, left axis).

Figure 4: A background-subtracted image with the equilibrium separatrix (red dashed line) and region of interest used to estimate the toroidal mode number (green solid line).

Figure 5: Histogram of the toroidal mode number in L-mode (blue, dashed) and H-mode (red) periods of a typical discharge.

Figure 6: Results from a blob tracking algorithm applied to the private flux region. The detected filaments are shown and their shape approximated by ellipses overlaid in yellow with major and minor axes indicated.

Figure 7: Measurements of the filament minor width, calculated by fitting image profiles along the minor axis of filaments to Gaussian curves to estimate the FWHM.

Figure 8: In a simplified model, a Gaussian light emissivity function (red) is line integrated to produce a curve (black) using an Abel transform method. The width (FWHM) of the line integrated profile is approximately twice that of the emissivity.

Figure 9: Measurements of the filament major width, calculated using interpolation to find the full-width at half-maximum of the light emitted by the filaments, taking into account line integration effects.

Figure 10: A series of profiles along an image showing the radial propagation of a filament and the apparent ejection of a secondary structure that propagates deeper into the PFR.

Figure 11: A typical ion saturation current profile measured by Langmuir probes at the inner divertor, plotted in terms of the average (solid, black) and standard deviation (red, hollow), showing the fluctuations in the ion current is offset from the peak in the mean value.

Figure 12: Profiles of ion saturation current at the inner divertor from Langmuir probes (black) and visible light emission from the camera (red), plotted in terms of normalised flux, $\psi_n$.

Figure 13: Profiles of light emitted from the inner target in the private flux region in two shots with different plasma current. The width of the average profile is shorter at higher $I_p$, but the fluctuations are unchanged.

Figure 14: Profiles of the average and standard deviation of ion saturation current measured at the outer divertor measured by Langmuir probes before (left) and after (right) a transition to H-mode.

Figure 15: Filaments generated in the PFR of the inner divertor leg would be expected to be strongly sheared by the X-point as they transit along field lines toward the outer divertor target.

Figure 16: Power spectrum of data recorded by the camera from a pixel viewing the private flux region of the inner leg in the H-mode period of a discharge. The power spectrum decays at frequencies greater than ~8kHz.

**Table Captions**

Table 1: Overview of the core plasma parameters scanned in the scaling study. *Defined as the ratio of $n_G$ ($10^{20}$m$^{-3}$)= $I_p$ (MA) / $a^2$ (m$^2$) [29] and the core line-average density.

# Figures

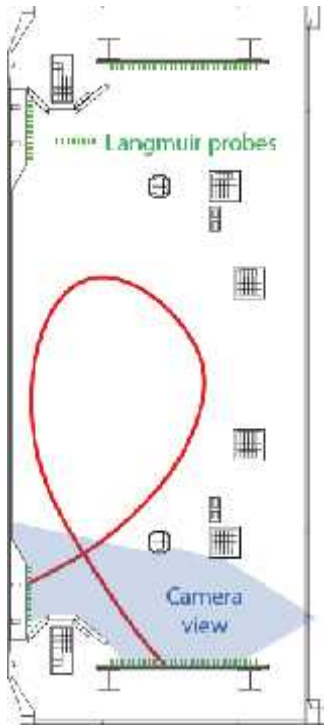

**Figure 1**

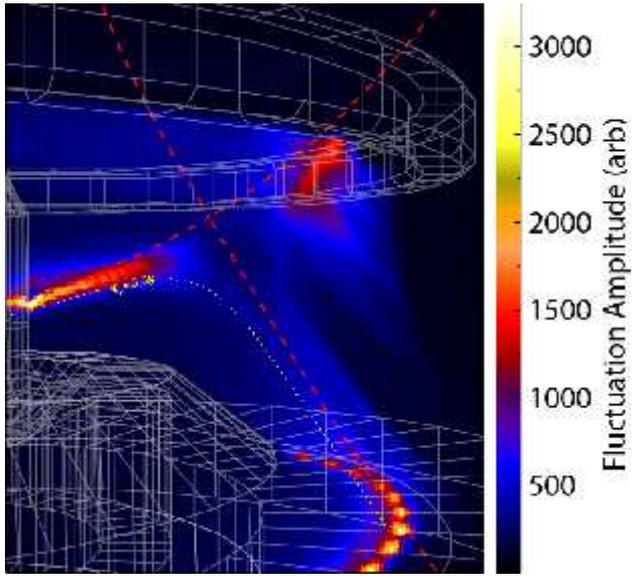

Figure 2

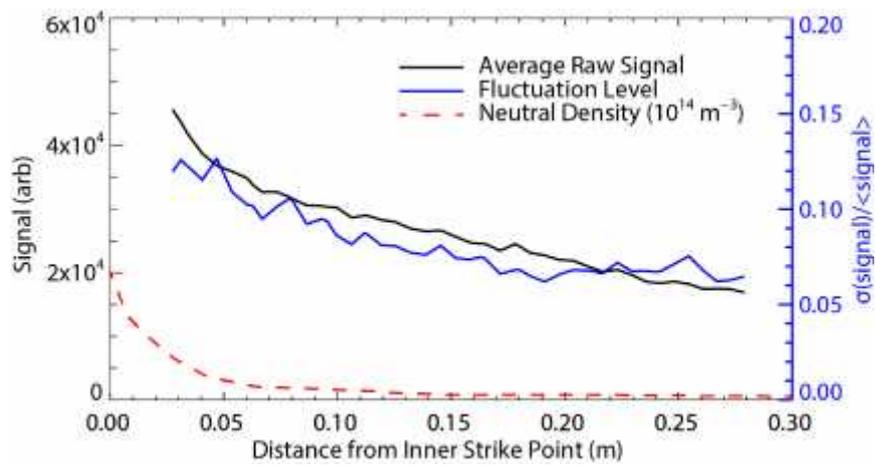

Figure 3

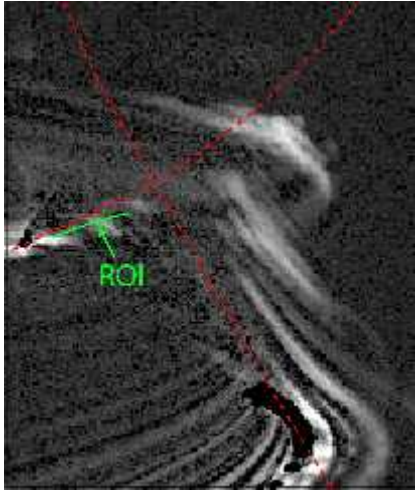

**Figure 4**

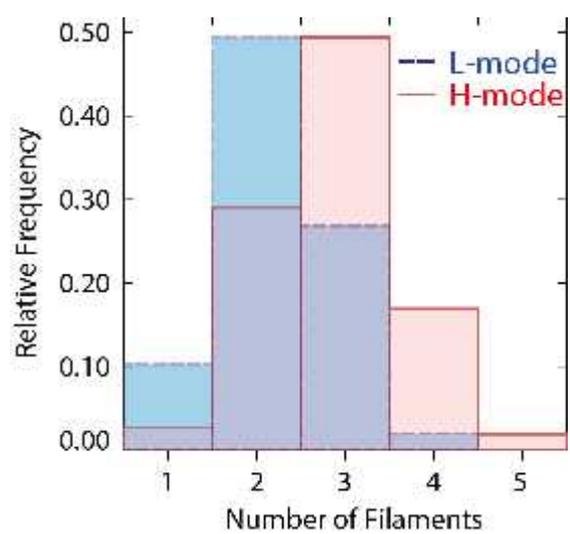

Figure 5

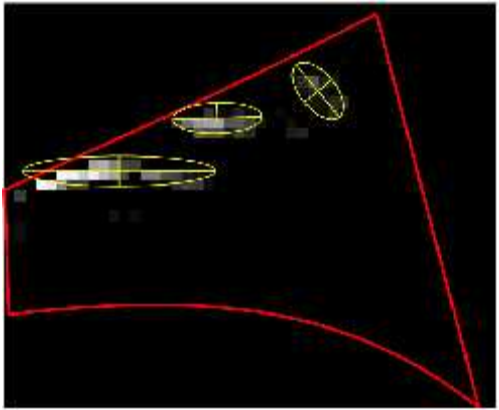

Figure 6

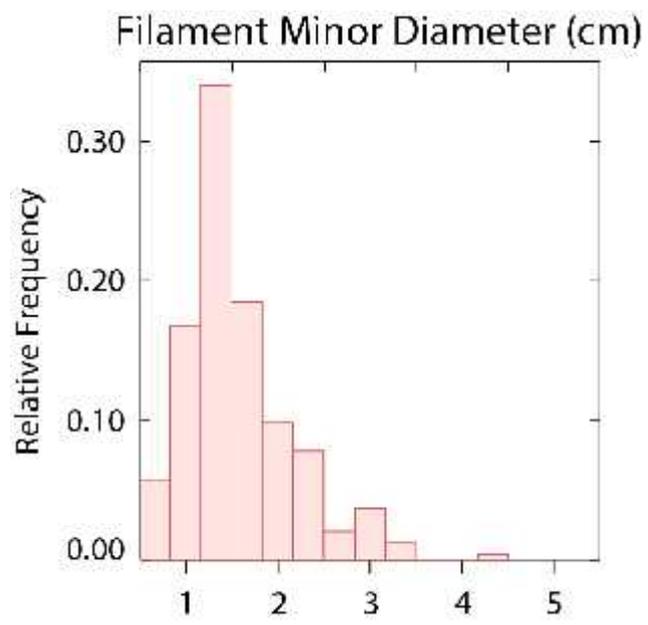

Figure 7

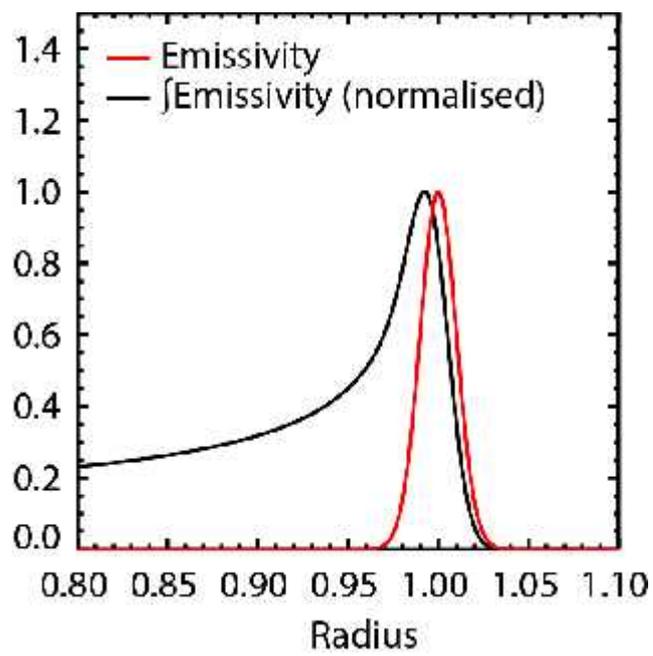

Figure 8

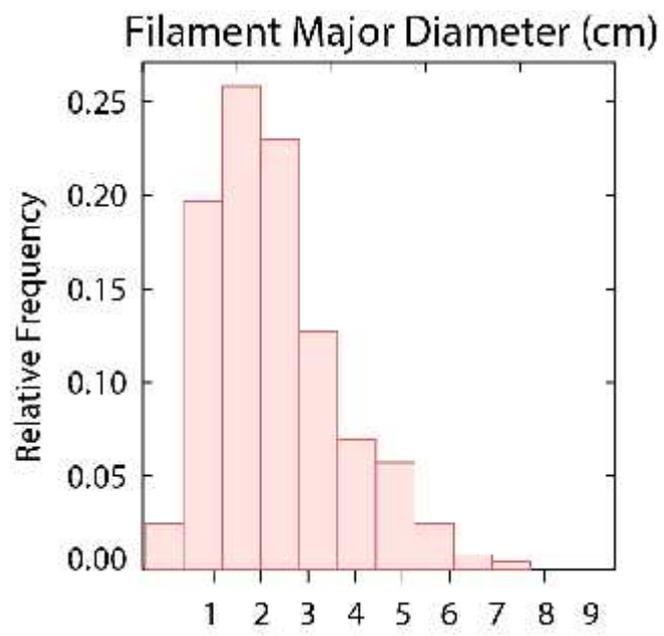

Figure 9

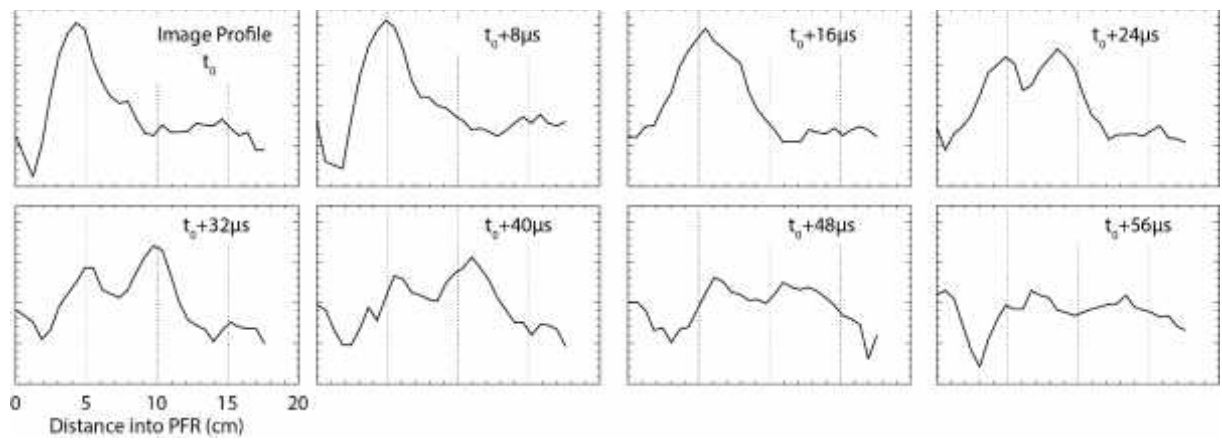

Figure 10

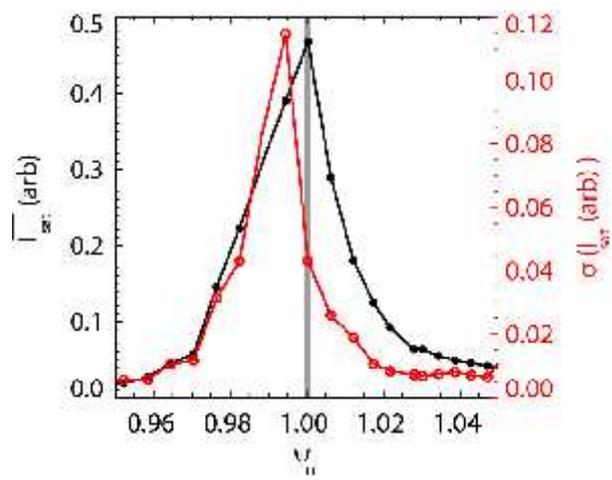

Figure 11

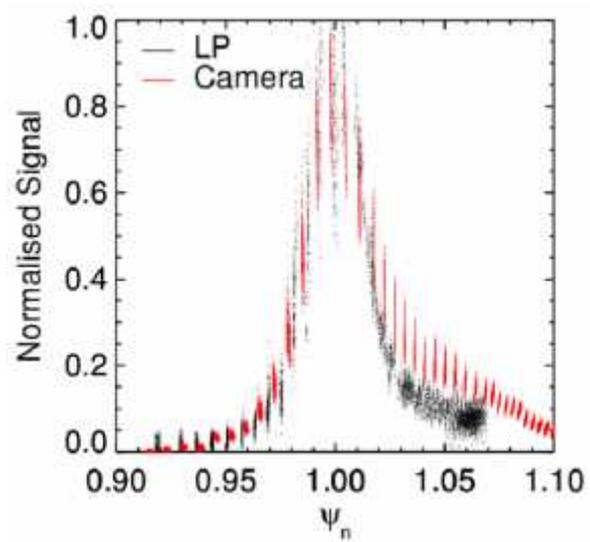

Figure 12

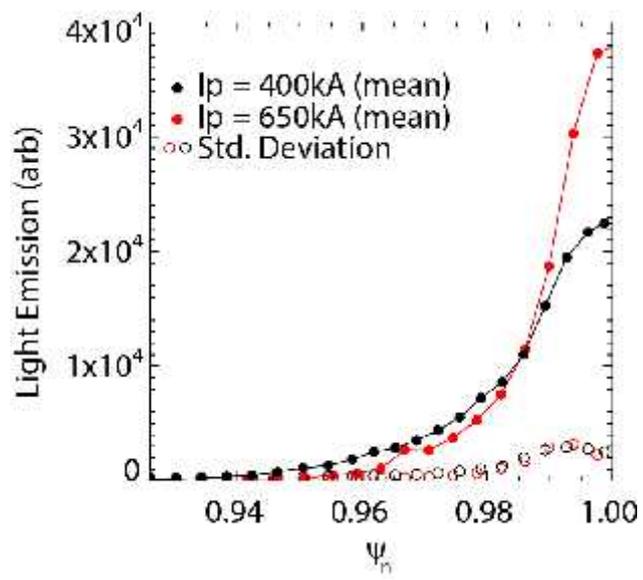

Figure 13

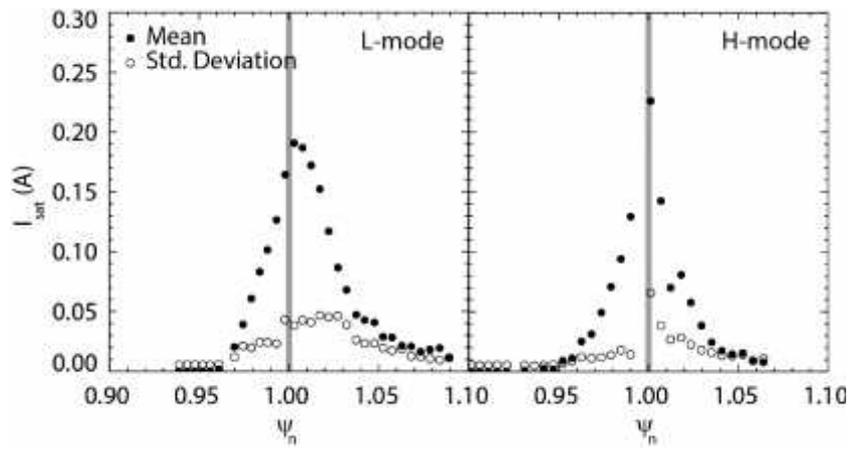

**Figure 14**

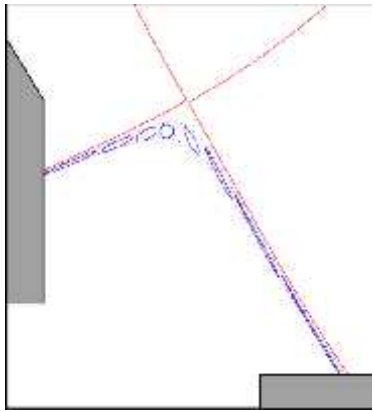

**Figure 15**

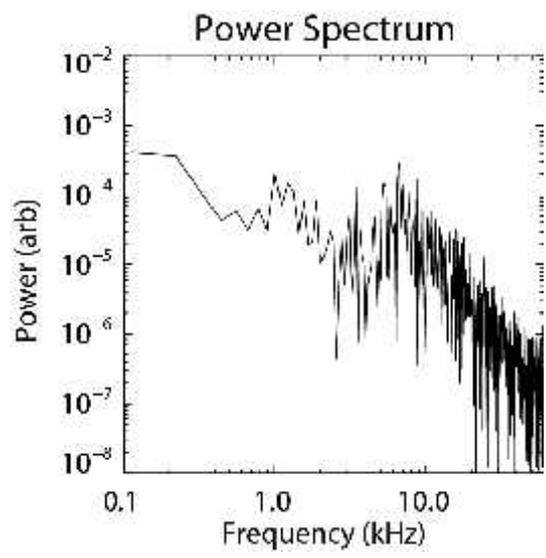

Figure 16

**Tables**

| Plasma Property | Scan Range |
|---|---|
| Core plasma density Greenwald fraction * | 0.27-0.49 |
| Auxiliary heating | 0MW-1.6MW |
| Magnetic topology | Single-null, double-null |
| Plasma current | 400kA-650kA |
| Toroidal field at magnetic axis | 0.585T |

Table 1